\documentclass[10pt,twocolumn,twoside]{IEEEtran}
\usepackage{ifpdf}
\usepackage{url}
\ifpdf
\else
\fi
\usepackage{pifont}
\usepackage{cite}
\usepackage{balance}
\usepackage{bm,comment,color}

\ifCLASSINFOpdf
  \usepackage[pdftex]{graphicx}
  \graphicspath{{../pdf/}{../jpeg/}}
  \DeclareGraphicsExtensions{.pdf,.jpeg,.png}
\else
  \usepackage[dvips]{graphicx}
  \graphicspath{{../eps/}}
  \DeclareGraphicsExtensions{.eps}
\fi
\usepackage{amsmath}
\usepackage{algpseudocode}
\algtext*{EndWhile}
\algtext*{EndIf}
\algtext*{EndFor}
\usepackage{algorithm}
\usepackage{array}
\usepackage{amsfonts} 
\usepackage{amssymb}
\usepackage{esint} 
\usepackage{units}
\usepackage{multirow}
\usepackage{comment}

\usepackage{amsmath}
\usepackage{stfloats} 

\usepackage{color}

\usepackage{standalone}

\usepackage{pgfplots}
\usepackage{tikz}
\usetikzlibrary{calc}
\makeatletter
\newcommand{\gettikzxy}[3]{%
  \tikz@scan@one@point\pgfutil@firstofone#1\relax
  \edef#2{\the\pgf@x}%
  \edef#3{\the\pgf@y}%
}
\usetikzlibrary{spy,backgrounds}
\pgfplotsset{compat=newest}
\usetikzlibrary{plotmarks}
\usetikzlibrary{arrows.meta}
\usepgfplotslibrary{patchplots}
\usepackage{grffile}
\newlength\fheight 
\newlength\fwidth 
\usepgfplotslibrary{fillbetween}

\usepackage{acronym}

\acrodef{6g}[6G]{the sixth generation}
\acrodef{aoa}[AOA]{angle-of-arrival}
\acrodef{bs}[BS]{base station}
\acrodef{bse}[BSE]{beam squint effect}
\acrodef{crb}[CRB]{Cram\'er-Rao bound}
\acrodef{elaa}[ELAA]{extremely large antenna array}
\acrodef{ff}[FF]{far-field}
\acrodef{las}[L\&S]{localization and sensing}
\acrodef{los}[LOS]{line-of-sight}
\acrodef{nf}[NF]{near-field}
\acrodef{nlos}[NLOS]{non-line-of-sight}
\acrodef{ofdm}[OFDM]{orthogonal frequency division multiplexing}
\acrodef{ris}[RIS]{reconfigurable intelligent surface}
\acrodef{sns}[SNS]{spatial non-stationarity}
\acrodef{swm}[SWM]{spherical wave model}
\acrodef{siso}[SISO]{single-input single-output}
\acrodef{ue}[UE]{user equipment}
\acrodef{dmimo}[D-MIMO]{distributed MIMO}

\usepackage{tabu,longtable}

\ifCLASSOPTIONcompsoc
 \usepackage[caption=false,font=normalsize,labelfont=sf,textfont=sf]{subfig}
\else
 \usepackage[caption=false,font=footnotesize]{subfig}
\fi
\usepackage{stfloats}
\usepackage[hidelinks]{hyperref}
\usepackage{xcolor}
\usepackage{graphicx}      
\newcommand{\circled}[1]{%
  \tikz[baseline=(char.base)]{
    \node[shape=circle, draw, inner sep=0.3pt, 
          minimum size=0.8em, align=center] (char) {\small #1};
  }
}

\usepackage{booktabs}

\usepackage{amsthm}

\usepackage{titlesec}
\titlespacing{\subsection}{0pt}{*0.5}{*0.3}
\titlespacing{\section}{0pt}{*0.5}{*0.3}
\renewcommand{\thesubsubsection}{\arabic{subsubsection})}
\makeatletter
\@addtoreset{subsubsection}{subsection}
\makeatother
\titleformat{\subsubsection}[runin]
  {\normalfont\normalsize\itshape}      %
  {\quad \thesubsubsection}          %
  {0.5em}                       %
  {}[:\hspace{0.5em}]                         %
\titlespacing{\subsubsection}{0pt}{*0.5}{1em} %

\ifCLASSINFOpdf
\else
\fi

\usepackage{amsmath}

\begin{document} 
\bstctlcite{IEEEexample:BSTcontrol}

\include{macros}
\title{Active Inference-Enabled Agentic Closed-Loop ISAC with Long-Horizon Planning}

\author{
Guangjin Pan\textsuperscript{*}, Zhuojun Tian\textsuperscript{†}, Mehdi Bennis\textsuperscript{‡}, Henk Wymeersch\textsuperscript{*}\\
\textsuperscript{*}Department of Electrical Engineering, Chalmers University of Technology, Sweden \\
\textsuperscript{†} Division of Information Science and Engineering, KTH Royal Institute of Technology, Sweden  \\
\textsuperscript{‡} Centre of Wireless Communications,  University of Oulu, Finland \\
\vspace{-8mm}
\thanks{This work was supported in part by the SNS JU project 6G-DISAC under the EU's Horizon Europe research and innovation Program under Grant Agreement No. 101139130, the Swedish Foundation for Strategic Research (SSF) (grant FUS21-0004, SAICOM), the ERANET CHIST-ERA Project MUSE-COM2, and in part by the Research Council of Finland (former Academy of Finland) Project Vision-Guided Wireless Communication. The computations were enabled by resources provided by the National Academic Infrastructure for Supercomputing in Sweden (NAISS), partially funded by the Swedish Research Council through grant agreement No. 2022-06725.}}

\maketitle

\begin{abstract}
Wireless agentic systems enable agents to autonomously perceive, reason, and act. However, existing works neglect the tight coupling between sensing and control in closed-loop integrated sensing and communication (ISAC) systems. In this paper, we propose an active inference (AIF)-driven wireless agentic system for closed-loop ISAC, which jointly optimizes control and sensing resource allocation via backward--forward message passing on a factor graph. The AIF agent maintains a generative model as a digital twin by integrating a localization model for uncertainty-aware state inference and a localization channel knowledge map (CKM) for approximating observation quality during planning. Simulation results demonstrate that the AIF-enabled agent adaptively allocates sensing resources based on spatially varying channel conditions, achieving superior balance among tracking accuracy, control effort, and sensing resource consumption over baseline strategies.
\end{abstract}

\section{Introduction}

The emergence of wireless agentic systems, where intelligent agents autonomously perceive, reason, and act upon the physical environment, is reshaping the design of next-generation wireless systems~\cite{dev2025advanced}. A prominent application of this paradigm lies in integrated sensing and communication (ISAC)~\cite{pan2025semantic}, where base stations (BSs) sense targets such as UAVs~\cite{pan2025active}, vehicles~\cite{Wang_AutonomousDriving_2019}, or robots~\cite{Goal-Oriented_robot}, and simultaneously deliver control commands to guide their behavior. In such closed-loop ISAC systems, the wireless agentic system forms a perception--cognition--action loop: the BS perceives the target's state through channel measurements, the intelligent agent within the BS infers the state of the target, plans future actions, and the target executes the resulting control decisions. This paradigm enables the wireless network to evolve from a passive communication infrastructure into an autonomous decision-making system that actively interacts with and shapes its physical environment.

A fundamental challenge in closed-loop ISAC is the tight coupling between sensing and control. On the one hand, the quality of state estimation depends on the allocated sensing resources, which directly affects control performance. On the other hand, control decisions determine the target's future trajectory, influencing the channel conditions and thus the future sensing quality. However, existing works largely neglect this coupling. Plenty of research focuses on ISAC resource allocation aimed at improving sensing performance~\cite{li2024maximizing} or balancing sensing and communication~\cite{khalili2024efficient, zou2024energy}. Other studies incorporate sensing results into closed-loop control, but typically assume fixed sensing quality, thereby ignoring the impact of sensing resource allocation on control performance~\cite{duan2022distributed, jin2025co}.
Furthermore, wireless agentic systems require a principled framework that unifies perception, reasoning, and action selection under uncertainty. Reinforcement learning (RL)-based approaches can learn joint policies, but they operate as black-box optimizers that lack interpretability and do not provide explicit uncertainty quantification over the agent's internal state~\cite{lockwood2022review}. This makes it difficult to understand why a particular sensing configuration is chosen or how state uncertainty drives resource allocation decisions.

Active inference (AIF), rooted in the free energy principle from neuroscience~\cite{friston2010free}, offers an interpretable and uncertainty-aware framework for autonomous agents, where state inference and action planning emerge from minimizing a unified free-energy objective that naturally balances goal achievement and uncertainty reduction. In this paper, we propose an AIF-driven wireless agentic system for closed-loop ISAC that jointly optimizes control and sensing resource allocation via message passing on a factor graph. The AIF-driven agent maintains a generative model as a digital twin, incorporating a state-transition model, a pretrained localization model for inference, and a localization channel knowledge map for approximating observation quality during planning. Compared with our preliminary work~\cite{pan2025active}, we make two key improvements. First, we replace the Cramér-Rao lower bound (CRLB)-based observation covariance approximation with learned neural network models. The CRLB provides only a theoretical lower bound on localization error, which can be loose in practical scenarios with complex non-linear channel-to-position mappings, whereas the learned models capture the actual localization performance. Second, we extend the single-step myopic planning to long-horizon backward--forward message passing, enabling uncertainty-aware decision-making that anticipates the impact of current sensing and control actions on future performance.

\color{black}

\begin{figure}[tb]
    \centering
    \begin{tikzpicture} 
    \node (image) [anchor=south west]{\includegraphics[scale=0.5]{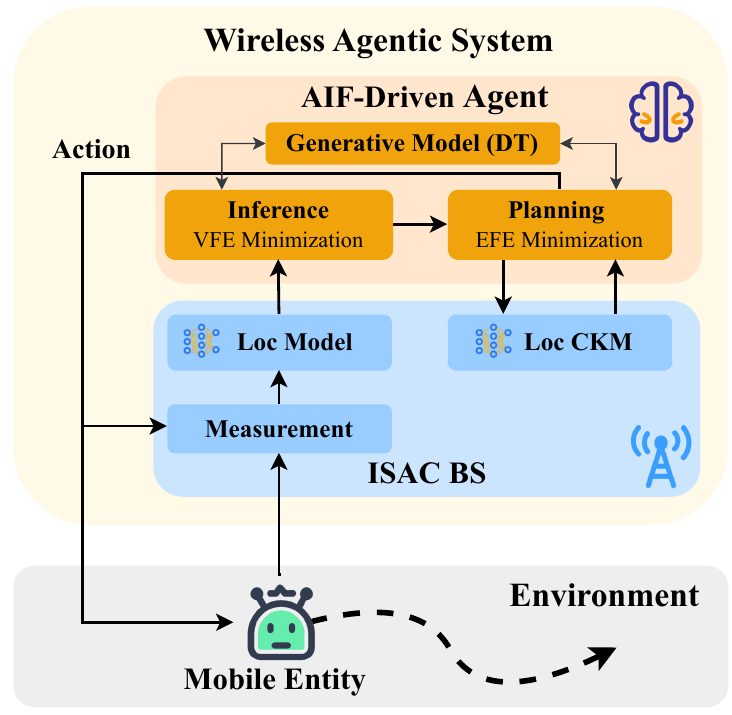}};        
    \gettikzxy{(image.north east)}{\ix}{\iy}; 
     \node[black, font=\scriptsize\bfseries] at (0.18*\ix, 0.435*\ix) {$k_{t+1}^*$};
     \node[black, font=\scriptsize\bfseries] at (0.18*\ix, 0.175*\ix) {$u_{t}^*$};

     \node[black, font=\scriptsize\bfseries] at (0.34*\ix, 0.445*\ix) {$H_{t}$};
     \node[black, font=\scriptsize\bfseries] at (0.323*\ix, 0.576*\ix) {$y_{t}, \hat{\Sigma}_t$};

     \node[black, font=\scriptsize\bfseries] at (0.57*\ix, 0.695*\ix) {$q(\!s_t\!)$};

     \node[black, font=\scriptsize\bfseries] at (0.89*\ix, 0.576*\ix) {$\hat{\Sigma}_t(k)$};
     \node[black, font=\scriptsize\bfseries] at (0.60*\ix, 0.576*\ix) {$l_x, l_y, k$};
    \end{tikzpicture}    
    \vspace{-5mm}
    \caption{AIF-enabled wireless agentic system for closed-loop ISAC.}
    \label{fig:system}
    \vspace{-5mm}
\end{figure}
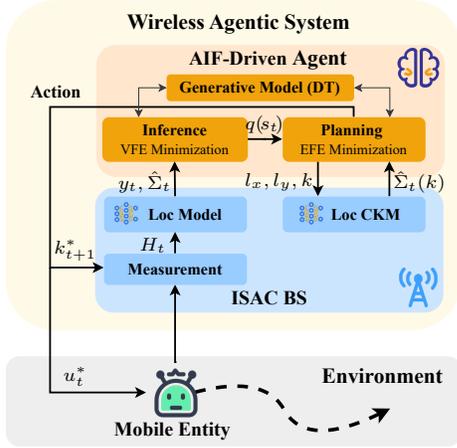

\section{System model}

As illustrated in Fig.~\ref{fig:system}, we consider a wireless agentic system for closed-loop sensing and control of a mobile entity (ME) (e.g., a UAV, robot, or autonomous vehicle). The system comprises an ISAC BS and an AIF-driven agent. The ISAC BS serves as the sensory interface, acquiring channel measurements and extracting position observations via a localization model. The AIF-driven agent serves as the cognitive core, maintaining an internal generative model that acts as a digital twin (DT) of the physical environment. At each time slot, the AIF-driven agent performs two operations: \emph{inference}, which updates the belief over the ME's state by minimizing the variational free energy (VFE), and \emph{planning}, which jointly optimizes the control input and sensing resource allocation by minimizing the expected free energy (EFE). The resulting actions are twofold: the control command is transmitted to the ME, and the sensing configuration is applied to the ISAC BS for the next measurement slot. This forms a closed-loop cycle driven by active inference.

\subsection{State Transition Model}

\label{subsec:state}
We consider the ME moving in a horizontal two-dimensional plane. The state at time $t$ is defined as $s_t \triangleq [l_{x,t}, l_{y,t}, v_{x,t}, v_{y,t}]^\top$, where $l_{x,t}$ and $l_{y,t}$ denote the position coordinates, and $v_{x,t}$ and $v_{y,t}$ denote the velocity components. With a sampling interval $\Delta t$, the state evolves according to the linear Gaussian model:
\begin{align}
    p(s_{t+1} \mid s_t, u_t)
    = \mathcal{N}(s_{t+1};\, A s_t + B u_t,\, Q),
    \label{eq:state_transition}
\end{align}
where $u_t \in \mathbb{R}^2$ is the acceleration command applied to the ME, $Q$ is the process noise covariance, corresponding to dynamics:
\begin{align}
    A =
    \begin{bmatrix}
        I_2 & \Delta t\, I_2 \\
        0   & I_2
    \end{bmatrix}, \quad
    B =
    \begin{bmatrix}
        \tfrac{1}{2}\Delta t^2\, I_2 \\
        \Delta t\, I_2
    \end{bmatrix}.
\end{align}

\subsection{Sensing and Observation Model}
\label{subsec:sensing}

The ME's state is partially observed through wireless channel measurements collected at the ISAC BS. At each time slot, the BS allocates $k_t \in \mathcal{K}$ subcarriers for localization, where $\mathcal{K}$ denotes the set of candidate subcarrier configurations. Each subcarrier has a spacing of $B_{\mathrm{sc}}$. Therefore, the effective sensing bandwidth at time slot $t$ is $k_t B_{\mathrm{sc}}$. We model the observation as the estimated position:
\begin{align}
    p(y_t \mid s_t, k_t) = \mathcal{N}(y_t;\, C s_t,\, \Sigma_t)
    \label{eq:observation}
\end{align}
where $C = [I_2,\, 0_{2 \times 2}]$ extracts the 2D position from $s_t$,
$y_t \in \mathbb{R}^2$ is the position observation, and
$\Sigma_t\in \mathbb{R}^{2 \times 2}$ is the observation
covariance. The covariance depends on the sensing resource $k_t$: allocating more subcarriers increases the effective bandwidth and reduces uncertainty $\Sigma_t$. The observation model~\eqref{eq:observation} is used in both the inference and planning stages, but the methods of obtaining $\Sigma_t$ are different as discussed below.

\subsubsection{Inference Stage: Localization Model}

During inference, the ISAC BS allocates $k_t$ subcarriers to acquire the channel frequency response $H_t$, which depends on the ME's state $s_t$ and the sensing configuration $k_t$. A pretrained localization model then maps the channel measurement to the observation and its covariance:
\begin{align}
    y_t,\, \hat\Sigma_t^{\mathrm{loc}} = \mathcal{F}^{\mathrm{loc}}_{\theta}(H_t).
    \label{eq:loc_model}
\end{align}
Since the true channel $H_t$ is available at time $t$, both $y_t$ and $\hat{\Sigma}_t^{\mathrm{loc}}$ are obtained directly from the measurements, where $\hat{\Sigma}_t^{\mathrm{loc}}$ serves as the estimate of $\Sigma_t$ in~\eqref{eq:observation}.

\subsubsection{Planning Stage: Localization CKM Model}
\label{subsubsec:cov_model}

During planning, the AIF-driven agent evaluates the observation quality at future positions under candidate sensing resource allocations, for which no real channel measurements exist. To bridge this gap, inspired by digital twin techniques, we proposed to incorporate a localization channel knowledge map (CKM) in the generative model, which predicts the expected observation covariance from position and sensing resource:
\begin{align}
    \hat{\Sigma}^{\mathrm{ckm}}(l_x, l_y,k) = \mathcal{F}^{\mathrm{ckm}}_{\phi}(
    l_x, l_y, k),
    \label{eq:cov_pred}
\end{align}
where $(l_x, l_y)$ is a queried position, $k$ is the candidate number of subcarriers, and $\phi$ denotes the model parameters. The predicted $\hat{\Sigma}^{\mathrm{ckm}}(l_x, l_y,k)$ serves as a surrogate for $\Sigma_t$ in~\eqref{eq:observation}, enabling the planning stage to approximate the observation model at future time steps. Since the planning stage needs to anticipate the localization accuracy
of $\mathcal{F}^{\mathrm{loc}}_{\theta}$ at future positions, $\mathcal{F}^{\mathrm{ckm}}_{\phi}$
is trained on the outputs of $\mathcal{F}^{\mathrm{loc}}_{\theta}$ to capture
its position- and subcarrier-dependent performance characteristics.

\subsection{Action Model and Objective}
\label{subsec:action}

The AIF-driven agent jointly determines the control input and the sensing resource allocation at each time slot. The action is $a_t = (u_t, k_{t+1})$, where $u_t$ drives the ME's state evolution through~\eqref{eq:state_transition}, and $k_{t+1}$ configures the ISAC BS for the next observation through~\eqref{eq:observation}.

The agent selects $a_t$ to balance control performance,
state-estimation accuracy, and sensing-resource consumption via the following cost terms:
\begin{itemize}
    \item \textbf{Observation cost:} Given the desired
    trajectory $s_t^{\mathrm{desired}}$ with
    $y_t^{\mathrm{desired}} = C s_t^{\mathrm{desired}}$,
    $J_t^{\mathrm{est}} = \tfrac{1}{2}
    (y_t - y_t^{\mathrm{desired}})^\top Q_{\mathrm{goal}}\,
    (y_t - y_t^{\mathrm{desired}})$,
    where $Q_{\mathrm{goal}} \succeq 0$ is a weighting matrix.
    \item \textbf{Control cost:}
    $J_t^{\mathrm{ctrl}} = \tfrac{1}{2}\,
    u_t^\top R_{\mathrm{goal}}\, u_t$,
    where $R_{\mathrm{goal}} \succeq 0$ penalizes the control
    energy to prevent excessive maneuvering.
    \item \textbf{Sensing-resource cost:}
    $J_t^{\mathrm{sens}} = \tfrac{1}{2}\,
    \alpha_{\mathrm{goal}}\, k_t^2$,
    where $\alpha_{\mathrm{goal}} \ge 0$ controls the tradeoff
    between sensing accuracy and resource consumption.
\end{itemize}
We consider long-term planning over a horizon of $T$ steps. The overall cost from the current time $t$ to the horizon end is
\begin{align}
    J = \sum_{\tau=t}^{t+T-1}  ( J_\tau^{\mathrm{ctrl}}
    + J_{\tau+1}^{\mathrm{est}}
    + J_{\tau+1}^{\mathrm{sens}}  ).
    \label{eq:objective}
\end{align}

\subsection{Problem Formulation via Active Inference}
\label{subsec:aif}

In active inference, the cost~\eqref{eq:objective} defines the agent's goal prior, a distribution encoding preferences over future observations and actions. The generative model maintained by the AIF-driven agent can be factorized as
\begin{align}
    &\! P_g(s_{1:t+T}, y_{1:t+T} \! \mid \! a_{1:t+T-1})
    = \,P_{g,1}(s_{1:t}, y_{1:t} \mid a_{1:t-1})
    \nonumber\\
    & \qquad \qquad \qquad  \times P_{g,2}(s_{t+1:t+T}, y_{t+1:t+T} \mid s_t, a_{t:t+T-1})
    \nonumber\\
    &\qquad \qquad \qquad \times \tilde{P}(y_{t+1:t+T}, u_{t:t+T-1}, k_{t+1:t+T}),
    \label{eq:gm_overall}
\end{align}
where
$P_{g,1} = \prod_{\tau=1}^{t-1}
p(s_{\tau+1} \! \mid \!  s_\tau, u_\tau)\,
p(y_{\tau+1} \!  \mid \!  s_{\tau+1}, k_{\tau+1})$ is the past generative factor, $P_{g,2} = \prod_{\tau=t}^{t+T-1}
p(s_{\tau+1} \mid s_\tau, u_\tau)\,
p(y_{\tau+1} \mid s_{\tau+1}, k_{\tau+1})$ characterizes the future state dynamics and observations, and $\tilde{P} = \prod_{\tau=t}^{t+T-1} \tilde{p}(y_{\tau+1}, u_\tau, k_{\tau+1})$ is the goal prior
with per-step factor
\begin{align}
    \tilde{p}(y_{\tau+1}, u_\tau, k_{\tau+1})
    = \frac{1}{Z}\exp (
    - J_\tau^{\mathrm{ctrl}}
    - J_{\tau+1}^{\mathrm{est}}
    - J_{\tau+1}^{\mathrm{sens}} ).
    \label{eq:goal_prior}
\end{align}
\color{black}
The AIF-driven agent operates through two stages:

\textbf{Inference stage.}
At time $t$, minimize VFE to update the belief:
\begin{align}
    \mathcal{F}_1 \!  =  \! \mathbb{E}_{q(s_{1:t} \mid y_{1:t})}\! [
    \log q(s_{1:t})
     \! - \! \log  \! P_{g,1}(s_{1:t}, y_{1:t}  \! \mid  \! a_{1:t-1})
     ].
    \label{eq:vfe_inference}
\end{align}


\textbf{Planning stage.} Minimize the EFE over the variational distributions $q(u_{t:t+T-1})$ and $q(k_{t+1:t+T})$:
\begin{align}
    \mathcal{G}_t
    &= \mathbb{E}_{q_t} \Bigg[
    \sum_{\tau=t}^{t+T-1} \Big(
    \log q(s_{\tau+1}, y_{\tau+1}, u_\tau, k_{\tau+1}
    \mid s_\tau)
    \nonumber\\
    & - \log P_{g,2}(s_{\tau+1}, y_{\tau+1}
    \mid s_\tau, u_\tau, k_{\tau+1}) \nonumber\\
    &- \log \tilde{p}(y_{\tau+1}, u_\tau, k_{\tau+1})
    \Big) \Bigg],
    \label{eq:efe}
\end{align}
where $q_t  \! \triangleq  \! q(s_{t+1:t+T}, y_{t+1:t+T}, 
u_{t:t+T-1}, k_{t+1:t+T} \! \mid  \! s_t)$  is the variational predictive distribution over future states, observations and actions. The optimal actions can be obtained as the modes of the
marginal variational factors, i.e., $u_\tau^* = \arg\max\, q(u_\tau)$ and $k_{\tau+1}^* = \arg\max\, q(k_{\tau+1})$.

\section{Active Inference-Based Joint Sensing and Control}
\label{sec:algorithm}

We solve~\eqref{eq:vfe_inference} and~\eqref{eq:efe} via message passing on the factor graph. At each time step $t$, the system operates as follows: (i) the ISAC BS collects $H_t$ using $k_t$ subcarriers; (ii) $\mathcal{F}^{\mathrm{loc}}_{\theta}$ extracts $(y_t, \hat{\Sigma}_t)$ and the AIF-driven agent updates the belief $q(s_t)$ via VFE minimization; (iii)  backward and forward message passing yield the beliefs $q(u_t)$ and $q(k_{t+1})$, where $\mathcal{F}^{\mathrm{ckm}}_{\phi}$ provides $\hat{\Sigma}(k)$ for planning; (iv) $u_t^*$ is transmitted to the ME and $k_{t+1}^*$ configures the BS for the next slot. In the following, we detail the inference and planning procedures, and describe the training of the localization model and CKM model.

\subsection{Inference Stage}
\label{subsec:inference}

The inference stage minimizes the VFE $\mathcal{F}_1$ in~\eqref{eq:vfe_inference} to obtain the posterior belief over $s_t$. Under the linear-Gaussian model, this admits a closed-form solution equivalent to Kalman filtering. At time $t$, the ISAC BS collects $H_t(s_t, k_t)$ and obtains $(y_t, \hat{\Sigma}_t)$ via~\eqref{eq:loc_model}. Given the belief $q(s_{t-1}) = \mathcal{N}(s_{t-1};\, \hat{m}_{t-1}^s,\, \hat{P}_{t-1}^s)$ and control $u_{t-1}$, the prediction step gives $\hat{m}_{t|t-1}^s = A\hat{m}_{t-1}^s + B u_{t-1}$ and $\hat{P}_{t|t-1}^s = A\hat{P}_{t-1}^s A^\top + Q$. Combining the transition and observation messages yields the posterior $q(s_t) = \mathcal{N}(s_t;\, \hat{m}_t^s,\, \hat{P}_t^s)$ with
\begin{align}
    \hat{P}_t^s &=  (
     (\hat{P}_{t|t-1}^s )^{-1}
    + C^\top (\hat\Sigma_t^{\mathrm{loc}})^{-1} C
     )^{-1},
    \label{eq:kf_update_cov}\\
    \hat{m}_t^s &= \hat{P}_t^s  (
    C^\top (\hat\Sigma_t^{\mathrm{loc}})^{-1}\, y_t
    +  (\hat{P}_{t|t-1}^s )^{-1}
    \hat{m}_{t|t-1}^s  ).
    \label{eq:kf_update_mean}
\end{align}
Here, $u_{t-1}$ and $k_t$ are determined by the planning stage at time $t\!-\!1$, Coupling planning and inference in a closed loop.

\begin{figure}[tb]
    \centering
    \includegraphics[scale=0.35]{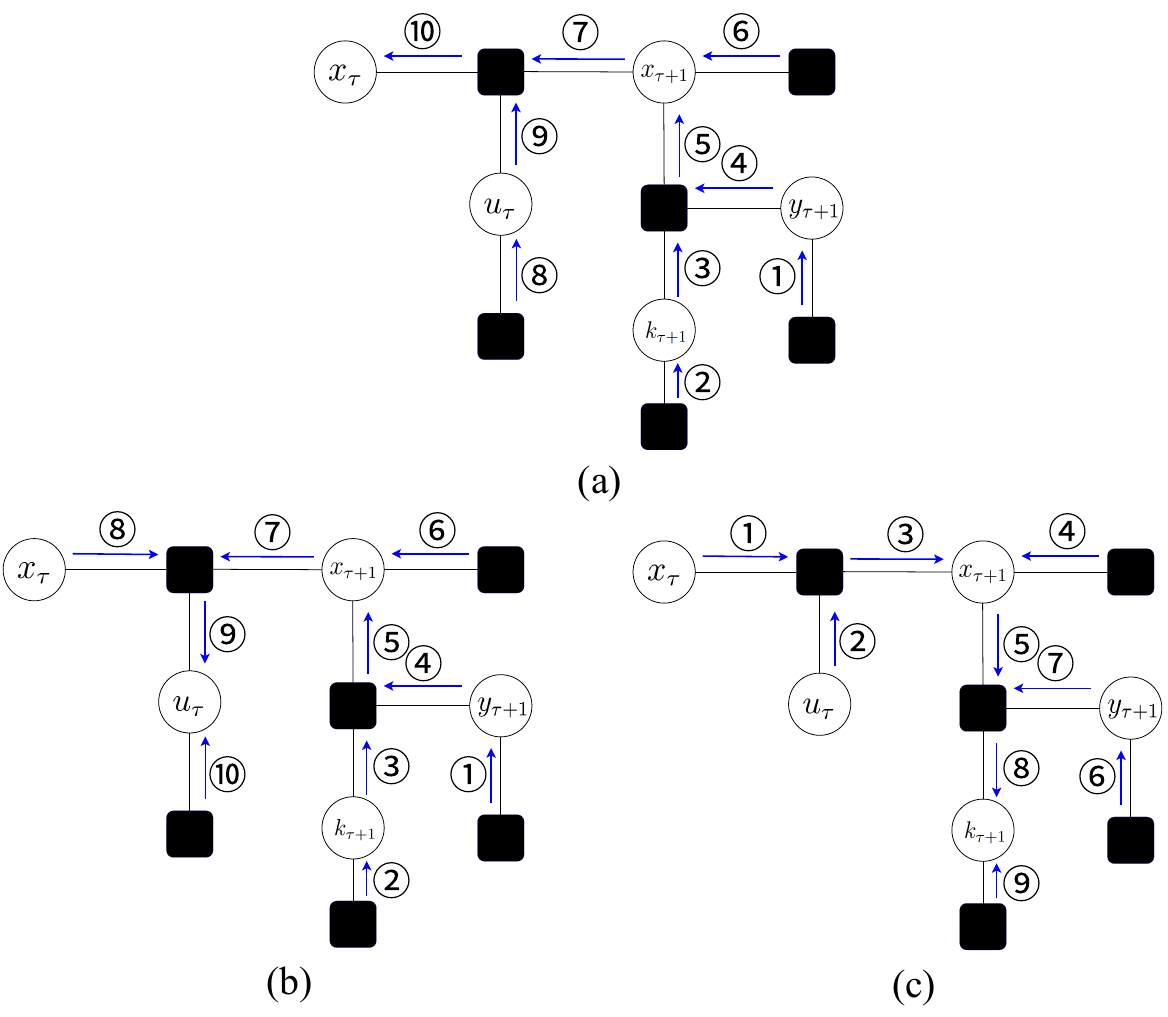}
    \vspace{-3mm}
        \caption{Factor graph for the planning stage: (a) backward pass for computing backward beliefs, (b) forward pass for control $u_\tau^*$, and (c) forward pass for sensing resource allocation $k_{\tau+1}^*$.}
    \label{fig:factor_graph}
    \vspace{-5mm}
\end{figure}

\subsection{Planning Stage}
\label{subsec:planning}

The planning stage minimizes the EFE in~\eqref{eq:efe} to determine the control $u_\tau$ and sensing allocation $k_{\tau+1}$ over a planning horizon from $t$ to $t+T$. We solve this via belief propagation on the factor graph, which is decomposed into a backward pass (encoding future preferences) and a forward pass (extracting optimal actions). The factor graph is illustrated in Fig.~\ref{fig:factor_graph}. The goal prior in~\eqref{eq:goal_prior} factorizes into
\begin{align}
    p_u(u_\tau) &= \mathcal{N}(u_\tau;\, 0,\,
    R_{\mathrm{goal}}^{-1}),
    \label{eq:prior_u} \\
    p_y(y_{\tau+1}) &= \mathcal{N}(y_{\tau+1};\,
    y_{\tau+1}^{\mathrm{desired}},\, Q_{\mathrm{goal}}^{-1}),
    \label{eq:prior_y} \\
    p_k(k_{\tau+1}) &= \mathcal{N}(k_{\tau+1};\, 0,\,
    \alpha_{\mathrm{goal}}^{-1}).
    \label{eq:prior_k}
\end{align}
In the planning stage, the observation covariance is replaced by
the predicted covariance $\hat{\Sigma}^{\mathrm{ckm}}(l_x, l_y, k)$ from~\eqref{eq:cov_pred}. For brevity, we write
$\hat{\Sigma}^{\mathrm{ckm}}(k_\tau) \triangleq
\mathcal{F}^{\mathrm{ckm}}_{\phi}(
l_{x,\tau}^{\mathrm{desired}},\,
l_{y,\tau}^{\mathrm{desired}},\, k_\tau)$,
where the position $(l_{x,\tau}^{\mathrm{desired}},l_{y,\tau}^{\mathrm{desired}})$ is taken from $s_\tau^{\mathrm{desired}}$
as a proxy for the real position at time $\tau$.

\subsubsection{Backward Pass}
\label{subsubsec:backwd_msg}

As shown in Fig.~\ref{fig:factor_graph} (a), the backward message from the observation factor to $s_{\tau+1}$ is obtained by marginalizing $y_{\tau+1}$ and $k_{\tau+1}$ over the observation likelihood, observation goal prior, and sensing cost prior (combining \text{\circled{1}}-- \text{\circled{4}} to obtain \text{\circled{5}}):
\begin{align}
    \mu_{\text{\circled{5}}} (s_{\tau+1})
    &\!\! \propto \!\! \sum_{k_{\tau+1} \!  \in  \mathcal{K}} \! p_k \! (k_{\tau+1})
    \! \int \!
    \mathcal{N}(y_{\tau+1};\, Cs_{\tau+1},\,
    \hat{\Sigma}^{\mathrm{ckm}}(k_{\tau+1})))
    \nonumber\\
    &\quad \times
    \mathcal{N}(y_{\tau+1};\,
    y_{\tau+1}^{\mathrm{desired}},\,
    Q_{\mathrm{goal}}^{-1})\,
    \mathrm{d}y_{\tau+1}.
    \label{eq:obs_msg_sum}
\end{align}

For each $k_{\tau+1}$, the integral over $y_{\tau+1}$ yields $\mathcal{N}(y_{\tau+1}^{\mathrm{desired}};\, Cs_{\tau+1},\, \Lambda_k)$ with $\Lambda_k \triangleq Q_{\mathrm{goal}}^{-1} + \hat{\Sigma}^{\mathrm{ckm}}(k_{\tau+1}))$. To extend to the full state dimension, we introduce a diffuse prior $\sigma^2 \gg 1$ on the velocity components, giving $\mathcal{N}(s_{\tau+1};\, \bar{\mu},\, \bar{\Sigma}_k)$ with $\bar{\mu} =
     [\begin{smallmatrix} y_{\tau+1}^{\mathrm{desired}} \\ 0
    \end{smallmatrix} ]$ and $
    \bar{\Sigma}_k = \mathrm{diag}(\Lambda_k,\sigma^2 I_2)$.
Approximating the mixture over $\mathcal{K}$ by moment matching yields
$\mu_{\text{\circled{5}}}(s_{\tau+1}) \propto
\mathcal{N}(s_{\tau+1};\, \bar{\mu},\, \bar{\Sigma}^*)$
with
$\bar{\Sigma}^* =  [\begin{smallmatrix}
\sum_{k \in \mathcal{K}} w_k \Lambda_k & 0 \\
0 & \sigma^2 I_2
\end{smallmatrix} ]$
and
$w_k = p_k(k) / \sum_{k' \in \mathcal{K}} p_k(k')$.
The observation message \text{\circled{5}} is then combined with the
backward belief, i.e., \text{\circled{6}}, $\mathcal{N}(s_{\tau+1};\, \hat{m}_{\tau+1}^{s,\mathrm{bw}},\, \hat{P}_{\tau+1}^{s,\mathrm{bw}})$, propagated from time
$\tau\!+\!2$, yielding $\mu_{\text{\circled{7}}}(s_{\tau+1})
\propto \mathcal{N}(s_{\tau+1};\,
\hat{m}_{\tau+1}^{s,\mathrm{tot}},\,
\hat{P}_{\tau+1}^{s,\mathrm{tot}})$ with
$  \hat{P}_{\tau+1}^{s,\mathrm{tot}}\!\!
    =\!\!  ( \!\!
     (\bar{\Sigma}^* )^{\!\!-1}\!\!\!\!
    + \!\! (\hat{P}_{\tau+1}^{s,\mathrm{bw}} )^{\!\!-1}  )^{-1}$ and $
    \hat{m}_{\tau+1}^{s,\mathrm{tot}}
    \!\! = \!\!\hat{P}_{\tau+1}^{s,\mathrm{tot}}  (\!\!
     (\bar{\Sigma}^* )^{-1} \bar{\mu}
    \!\! + \!\! (\hat{P}_{\tau+1}^{s,\mathrm{bw}} )^{-1}\!\!
    \hat{m}_{\tau+1}^{s,\mathrm{bw}} \!\!  ).$
Marginalizing $s_{\tau+1}$ and $u_\tau$ through the transition factor and control prior based on \text{\circled{7}} and \text{\circled{9}}, we obtain the backward belief
$\mu_{\text{\circled{10}}}(s_\tau) \propto
\mathcal{N}(s_\tau;\, \hat{m}_\tau^{s,\mathrm{bw}},\,
\hat{P}_\tau^{s,\mathrm{bw}})$ with $
    \hat{P}_\tau^{s,\mathrm{bw}}
    = (A^\top S_\tau^{-1} A)^{-1} $ and $
    \hat{m}_\tau^{s,\mathrm{bw}}
    = \hat{P}_\tau^{s,\mathrm{bw}}\,
    A^\top S_\tau^{-1}\, \hat{m}_{\tau+1}^{s,\mathrm{tot}}$, where $ S_\tau = \hat{P}_{\tau+1}^{s,\mathrm{tot}} + Q
    + B R_{\mathrm{goal}}^{-1} B^\top$.
At the terminal time $\tau = t+T$, the backward belief is initialized
as uninformative: $\hat{m}_{t+T}^{s,\mathrm{bw}} = 0$ and
$\hat{P}_{t+T}^{s,\mathrm{bw}} = \sigma_0^2 I$ with
$\sigma_0^2 \gg 1$. The recursion then proceeds
from $\tau = {t+T}$ down to $\tau = t$.

\subsubsection{Forward Pass}
\label{subsubsec:forward}

The forward pass starts from the posterior belief $q(s_t) = \mathcal{N}(s_t;\, \hat{m}_t^s,\, \hat{P}_t^s)$ obtained in the inference stage. At each step $\tau$, the control $u_\tau$ is first optimized, and then the sensing allocation $k_{\tau+1}$ is determined given $u_\tau^*$. Since the system operates in a receding-horizon fashion, only the first-step actions $(u_t^*, k_{t+1}^*)$ are executed, and the planning is repeated at the next time slot with updated belief.

\textbf{Optimal control:}
As shown in Fig.~\ref{fig:factor_graph} (b), the forward message at $s_\tau$ is $\mu_{\text{\circled{8}}}(s_\tau) \propto q(s_\tau) = \mathcal{N}(s_\tau;\, \hat{m}_\tau^{s,f},\, \hat{P}_\tau^{s,f})$, where $q(s_t)$ is initialized from the inference-stage posterior~\eqref{eq:kf_update_cov}--\eqref{eq:kf_update_mean} and propagated forward for $\tau > t$.
The backward message \text{\circled{7}} at $s_{\tau+1}$ is $\mathcal{N}(s_{\tau+1};\, \hat{m}_{\tau+1}^{s,\mathrm{tot}},\, \hat{P}_{\tau+1}^{s,\mathrm{tot}})$. Fusing \text{\circled{8}} and \text{\circled{7}} through the transition factor by marginalizing $s_{\tau+1}$ yields the message \text{\circled{10}} toward $u_\tau$. Multiplying \text{\circled{10}} with the control prior \text{\circled{9}}, $p_u(u_\tau) = \mathcal{N}(u_\tau;\, 0,\, R_{\mathrm{goal}}^{-1})$, the posterior over $u_\tau$ is $q(u_\tau)
    = \mathcal{N}(u_\tau;\, \hat{m}_\tau^u,\, P_\tau^u),
    \label{eq:q_u} $
where
$
\hat{P}_\tau^{s,\mathrm{pred}} =  A\hat{P}_\tau^{s,f} A^\top + Q$, $ 
    D_\tau = \hat{P}_\tau^{s,\mathrm{pred}}
    + \hat{P}_{\tau+1}^{s,\mathrm{tot}}$,$
    P_\tau^u = (B^\top D_\tau^{-1} B
    + R_{\mathrm{goal}})^{-1}$, and $ hat{m}_\tau^u = P_\tau^u\, B^\top D_\tau^{-1}\,
    (\hat{m}_{\tau+1}^{s,\mathrm{tot}}
    - A\hat{m}_\tau^{s,f}).
$
The optimal control is the mode of $q(u_\tau)$, i.e., $u_\tau^* = \hat{m}_\tau^u$.

\textbf{Sensing resource allocation:}
As shown in Fig.~\ref{fig:factor_graph} (c), given $u_\tau^*$, we first fuse the forward messages \text{\circled{1}}-- \text{\circled{3}}, which encode the state estimate and control action, with the backward message \text{\circled{4}}, which encodes future planning information, to obtain $\mu_{\text{\circled{5}}}(s_{\tau+1}) \propto
\mathcal{N}(s_{\tau+1};\, \hat{m}_{\tau+1}^{s,\mathrm{fuse}},\, \hat{P}_{\tau+1}^{s,\mathrm{fuse}})$ with
\begin{align}
    \!\hat{P}_{\tau+1}^{s,\mathrm{fuse}}
    &=  (
     (\hat{P}_\tau^{s,\mathrm{pred}} )^{-1}
    +  (\hat{P}_{\tau+1}^{s,\mathrm{bw}} )^{-1}
     )^{-1},
    \label{eq:P_fuse}\\
    \!\hat{m}_{\tau+1}^{s,\mathrm{fuse}}
    &= \hat{P}_{\tau+1}^{s,\mathrm{fuse}}    (
      (\hat{P}_\tau^{s,\mathrm{pred}} )^{-1}  \!
    \hat{s}_{\tau+1}^{\mathrm{pred}}
     \! + \!  (\hat{P}_{\tau+1}^{s,\mathrm{bw}} )^{-1}
      \hat{m}_{\tau+1}^{s,\mathrm{bw}}  ),
    \label{eq:m_fuse}
\end{align}
where $\hat{s}_{\tau+1}^{\mathrm{pred}} = A\hat{m}_\tau^{s,f} + B u_\tau^*$. Passing \text{\circled{5}} through the observation factor and the observation goal prior \text{\circled{6}} by marginalizing over $s_{\tau+1}$ and $y_{\tau+1}$ yields $ \mu_{\text{\circled{8}}}(k_{\tau+1})
    \propto
    \mathcal{N}\! (
    C\, \hat{m}_{\tau+1}^{s,\mathrm{fuse}};\,
    y_{\tau+1}^{\mathrm{desired}},\, V_k  )$,
where $V_k = Q_{\mathrm{goal}}^{-1}  + \hat{\Sigma}^{\mathrm{ckm}}(k_{\tau+1}) + C\, \hat{P}_{\tau+1}^{s,\mathrm{fuse}}\, C^\top $.

Multiplying \text{\circled{8}} with the sensing cost prior
\text{\circled{9}}, the posterior over $k_{\tau+1}$ is $q(k_{\tau+1})
    \propto p_k(k_{\tau+1}) \cdot
    \mu_{\text{\circled{8}}}(k_{\tau+1})$,
and the optimal sensing allocation is
$k_{\tau+1}^* = \arg\max_{k \in \mathcal{K}}\, q(k_{\tau+1})$,
evaluated by enumeration since $|\mathcal{K}|$ is small.


\subsection{Model Training}
\label{subsec:training}

The proposed algorithm relies on two pre-trained components: the localization model $\mathcal{F}^{\mathrm{loc}}_{\theta}$ used in the inference stage, and the covariance prediction model $\mathcal{F}^{\mathrm{ckm}}_{\phi}$ used in the planning stage. Both are trained offline prior to deployment.

\subsubsection{Localization Model}

The localization model $\mathcal{F}^{\mathrm{loc}}_{\theta}$ employs a ResNet-34 \cite{pan2025ai} backbone as a shared feature extractor. The extracted features are fed into a position head and variance head, each consisting of a two-layer MLP (64 hidden units, 2 outputs). Given the channel $H_t$, the position head outputs the location estimate
$[\hat{l}_{x,t},\, \hat{l}_{y,t}]$, while the variance head
outputs $[\hat{\sigma}_{x,t}^{\mathrm{loc},2},\,
\hat{\sigma}_{y,t}^{\mathrm{loc},2}]$, yielding
$\hat{\Sigma}_t^{\mathrm{loc}} = \mathrm{diag}(
\hat{\sigma}_{x,t}^{\mathrm{loc},2},\,
\hat{\sigma}_{y,t}^{\mathrm{loc},2})$.

Training proceeds in two stages. In Stage~1, the backbone and position head are trained by minimizing the mean squared error (MSE) over the 2D position using data from all $k \in \mathcal{K}$. In Stage~2, the backbone is frozen and the $|\mathcal{K}|$ variance heads are trained independently by minimizing the negative log-likelihood (NLL) \cite{pinto2021uncertainty}:
\begin{align}
    \mathcal{L}_{\mathrm{var}}
    = \frac{1}{2N_{\text{batch}}} \sum_{n=1}^{N_{\text{batch}}} \sum_{d \in \{x,y\}}
    \! ( \log \hat{\sigma}_{d,n}^{\mathrm{loc},2}
    + \frac{(l_{d,n} - \hat{l}_{d,n})^2}{\hat{\sigma}_{d,n}^{\mathrm{loc},2}}
     ) \nonumber
\end{align}
where $N_{\text{batch}}$ is the batch size. We train a separate variance head for each $k$ rather than a
single shared head, because different subcarrier configurations
lead to different noise levels in the position observations. A shared head would be forced to average over these heterogeneous noise distributions, resulting in variance estimates that are neither accurate for small $k$ (underestimating uncertainty) nor for large $k$ (overestimating uncertainty). Separate heads allow each to specialize in the noise regime of its corresponding $k$, resulting in well-calibrated covariance estimates across all sensing configurations.

\subsubsection{Localization CKM Model}

During planning, future channels are unavailable, where $\mathcal{F}^{\mathrm{loc}}_{\theta}$ cannot be used to obtain $\Sigma(k)$. To solve this problem, we train a lightweight MLP $\mathcal{F}^{\mathrm{ckm}}_{\phi}(\cdot)$ with two hidden layers of 128 units each and a 2-dimensional output layer, which predicts the observation covariance from position and sensing configuration:
\begin{align}
    [\hat{\sigma}_x^{\mathrm{ckm},2},\,
    \hat{\sigma}_y^{\mathrm{ckm},2}]
    = \mathcal{F}^{\mathrm{ckm}}_{\phi}(l_x, l_y, k),
    \label{eq:cov_mlp}
\end{align}
with $\hat{\Sigma}^{\mathrm{ckm}}(l_x, l_y,k) = \mathrm{diag}(
\hat{\sigma}_x^{\mathrm{ckm},2},\,
\hat{\sigma}_y^{\mathrm{ckm},2})$. To generate training data, we randomly sample positions $(l_x, l_y)$ and subcarrier configurations $k \in \mathcal{K}$, simulate the corresponding channels, and apply $\mathcal{F}^{\mathrm{loc}}_{\theta}$ to obtain the predicted variance $\hat{\sigma}_{d}^2$ for each sample. The MLP is then trained by minimizing the MSE between its output and the variance predicted by $\mathcal{F}^{\mathrm{loc}}_{\theta}$:
\begin{align}
    \mathcal{L}_{\mathrm{ckm}}
    = \frac{1}{2N_{\text{batch}}} \sum_{n=1}^{N_{\text{batch}}} \sum_{d \in \{x,y\}}
     ( \log\hat{\sigma}_{d,n}^{\mathrm{ckm},2}
    - \log\hat{\sigma}_{d,n}^{\mathrm{loc},2}
     )^2.
    \label{eq:loss_ckm}
\end{align}

\section{Numerical Results}
\label{sec:results}

We consider the Sionna~\cite{sionna} street canyon scene. The BS
is at $(-32, 12, 35)$ with a $16$-element ULA at half-wavelength spacing, operating at $f_c = 3$ GHz with a transmit power of $23$ dBm and noise power spectral density of $-174$ dBm/Hz. The ME has a single isotropic antenna at $1.5$\,m height. The set of candidate subcarrier configurations is $\mathcal{K} = \{50, 100, 150, 200, 250, 300, 350, 400\}$, corresponding to effective sensing bandwidths from $12.5$~MHz to $100$~MHz.
The training dataset consists of $20{,}000$ randomly sampled ME positions generated via Sionna ray tracing with up to 3 interaction bounces and diffraction enabled. The dataset is split into $15{,}000$ samples for Stage~1 (location training) and $5{,}000$ samples for Stage~2 (variance training), with non-overlapping splits to prevent overfitting. Both models are trained with the Adam optimizer at a learning rate of $10^{-4}$ for 200 epochs.
For the closed-loop simulation, the ME follows a straight-line reference trajectory from $(-50, 0)$\,m to $(50, 0)$\,m at a constant velocity of $v_{\mathrm{ref}} = 1$\,m/s, yielding $N = 1000$ time steps with $\Delta t = 0.1$\,s. The process noise covariance is $Q = \sigma_w^2 I_4$ with $\sigma_w = 0.01$. AIF goal-prior parameters are $Q_{\mathrm{goal}} = \mathrm{diag}(0.5, 0.5)$, $R_{\mathrm{goal}} = \mathrm{diag}(0.1, 0.1)$, and $\alpha_{\mathrm{goal}} = 1 \times 10 ^{-6}$. The planning horizon is $T = 10$ unless otherwise stated. All results are averaged over 10 independent runs with different random seeds.

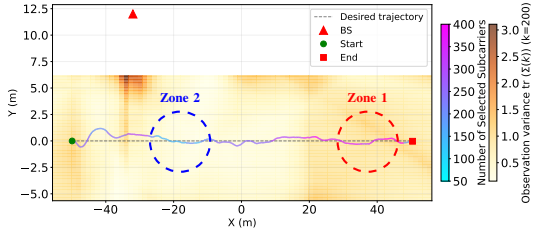
\begin{figure}[tb]
    \centering
    \begin{tikzpicture}
    \node (image) [anchor=south west]{\includegraphics[width=0.8\linewidth]{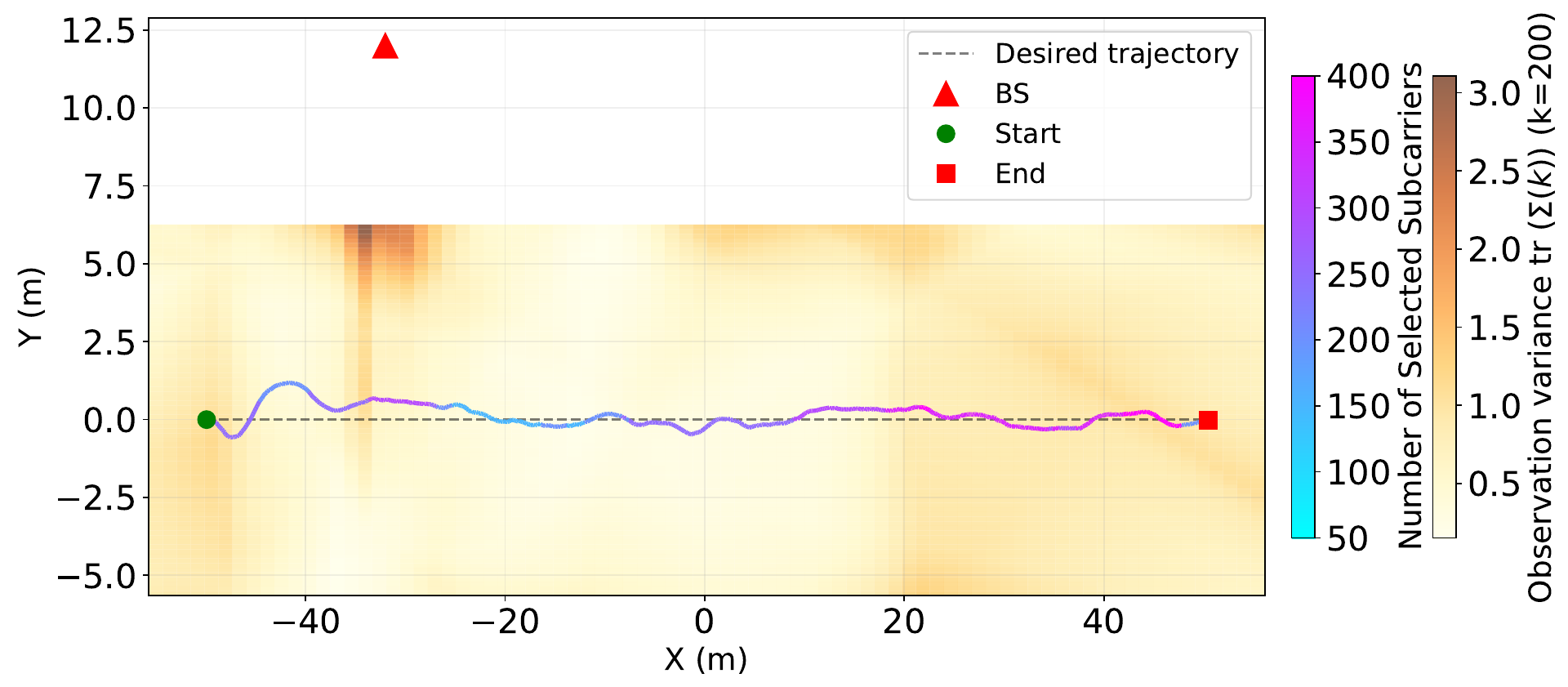}};        
    \gettikzxy{(image.north east)}{\ix}{\iy};   
            \draw[red, thick, dashed] (0.68*\ix, 0.18*\ix) circle (0.4);
            \node[red, font=\tiny\bfseries] at (0.68*\ix, 0.26*\ix) {Zone~1};
            \draw[blue, thick, dashed] (0.34*\ix, 0.18*\ix) circle (0.4);
            \node[blue, font=\tiny\bfseries] at (0.34*\ix, 0.26*\ix) {Zone~2};
        \end{tikzpicture}    
    \vspace{-4mm}
    \caption{AIF trajectory colored by the selected $k$, overlaid on the observation variance map. AIF allocates more subcarriers in high-variance regions (Zone~1) and fewer in low-variance regions (Zone~2).}
    \label{fig:traj}
    \vspace{-5mm}
\end{figure}

Fig.~\ref{fig:traj} shows the trajectory overlaid on the localization variance map (for $k=200$ subcarriers). The trajectory color encodes the subcarrier count $k$ selected by AIF at each time step. In Zone~1, where the predicted observation variance is high, AIF allocates more subcarriers to reduce localization uncertainty. In Zone~2, where the variance is low, AIF reduces the sensing resources. This confirms that the proposed method adaptively balances sensing accuracy and resource consumption based on the spatially varying channel quality.

Table~\ref{tab:comparison} compares the AIF method against two
baselines over 10 planning steps, averaged over 10 independent
runs: (i)~AIF control with sensing resources sampled from the
prior $p_k(k)$ (prior-$k$ + AIF-$u$), and (ii)~AIF sensing
allocation with greedy control that drives the next-step position
to the desired target (AIF-$k$ + greedy-$u$). The proposed AIF achieves the lowest total cost and the lowest average EFE, confirming that jointly optimizing control and sensing via AIF yields the best overall performance. Compared with prior-$k$ + AIF-$u$, AIF reduces the observation cost by $47\%$ through adaptive sensing allocation. The greedy-$u$ baseline
achieves the smallest observation cost but incurs an excessive control cost, as it ignores the control penalty entirely.

\begin{figure}[t]
\centering
\definecolor{color1}{RGB}{214,39,40}
\begin{tikzpicture}[font=\footnotesize]
\begin{axis}[%
    width=0.85\columnwidth,
    height=2.5cm,
    scale only axis,
    xmin=0, xmax=21,
    xtick={0,1,5,10,15,20},
    xmin=0, xmax=20,
    xlabel={Planning horizon length},
    ymin=0, ymax=1.8,
    ytick={0,0.6,1.2,1.8},
    ylabel={Average AIF Cost},
    xlabel style={font=\small},
    ylabel style={font=\small},
    tick label style={font=\small},
    axis background/.style={fill=white},
    xmajorgrids, ymajorgrids,
    grid style={dashed, gray!30},
    legend style={fill opacity=0.8, draw opacity=1, text opacity=1, at={(0.97,0.97)}, anchor=north east, font=\scriptsize, draw=white!15!black},
]
\addplot [color=color1, line width=1.0pt, mark=triangle*, mark options={solid, fill=color1, scale=1.2}]
  table[row sep=crcr]{%
1    1.029 \\
5    0.188  \\
10   0.151 \\
15   0.149  \\
20   0.149 \\
};
\addlegendentry{AIF (proposed)}

\addplot [color=blue, line width=1.0pt, mark=*, mark options={solid, fill=blue, scale=1.2}]
  table[row sep=crcr]{%
1    1.288 \\
5    0.315  \\
10   0.243 \\
15   0.248  \\
20   0.236 \\
};
\addlegendentry{Prior-$k$ + AIF-$u$}

\addplot [color=black, line width=1.0pt, mark=star, mark options={solid, fill=black, scale=1.2}]
  table[row sep=crcr]{%
1    1.598 \\
5    1.554  \\
10   1.478 \\
15   1.459  \\
20   1.463 \\
};
\addlegendentry{AIF-$k$ + Greedy-$u$}

\end{axis}
\end{tikzpicture}
\vspace{-4mm}
\caption{Average AIF cost under different planning horizon lengths.}
\label{fig:horizon}
\vspace{-4mm}
\end{figure}
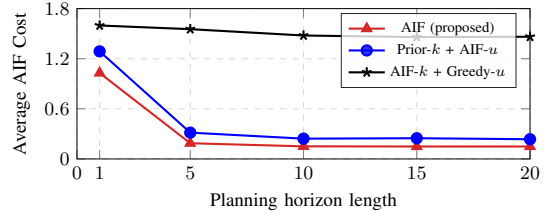

Fig.~\ref{fig:horizon} shows the average AIF cost versus planning horizon for all three methods. The proposed AIF achieves the lowest cost across all horizons, dropping by $85\%$ from horizon~$1$ to $10$ and saturating thereafter. The prior-$k$ + AIF-$u$ baseline follows a similar trend but converges to a higher cost, confirming the benefit of adaptive sensing allocation. AIF-$k$ + greedy-$u$ improves only marginally with longer horizons, as the greedy control limits the agent's ability to exploit long-horizon planning. These results confirm that jointly optimizing sensing and control with long-horizon planning is essential for closed-loop performance.

\begin{table}[t]
    \centering
    \caption{Cost Comparison of Different Strategies}
    \label{tab:comparison}
    \begin{tabular}{@{}lcccccc@{}}
        \toprule
        \textbf{Method} & $J^{\mathrm{est}}$ & $J^{\mathrm{ctrl}}$ & $J^{\mathrm{sens}}$ & $J$ & Ave. EFE \\
        \midrule
        AIF (proposed)          & 110.01 & \textbf{1.61}  & 39.05 & \textbf{150.67} & \textbf{$-$1.453} \\
        Prior-$k$ + AIF-$u$    & 206.55 & 3.70  & \textbf{32.77} & 243.02 & -0.695 \\
        AIF-$k$ + Greedy-$u$   & \textbf{82.12}  & 1357.73 & 37.97 & 1477.82 & -0.079 \\
        \bottomrule
    \end{tabular}
    \vspace{-4mm}
\end{table}

\color{black}

\section{Conclusion}
In this paper, we propose an AIF-driven wireless agentic system for closed-loop ISAC that jointly optimizes control and sensing resource allocation. The AIF agent incorporates a pretrained localization model for uncertainty-aware state inference and a CKM-based covariance prediction model for long-horizon planning. Simulation results confirm that the proposed method adaptively allocates sensing resources based on spatially varying channel conditions, achieving superior balance among tracking accuracy, control effort, and sensing resource consumption over baseline strategies.

\balance 
\bstctlcite{IEEEexample:BSTcontrol}
\bibliographystyle{IEEEtran}
\bibliography{IEEEabrv,ref}


\end{document}